\documentclass[useAMS,usenatbib]{mn2e}
\usepackage{graphicx}

\def\H0{{\rm ~km~s^{-1}~Mpc^{-1}}}

\def\la{\mathrel{\hbox{\rlap{\hbox{\lower4pt\hbox{$\sim$}}}{\raise2pt\hbox{$<$}}}}}
\def\ga{\mathrel{\hbox{\rlap{\hbox{\lower4pt\hbox{$\sim$}}}{\raise2pt\hbox{$>$}}}}}
\def\d25{D$_{25}$}

\title[A new constraint for gamma-ray burst progenitor mass]
{A new constraint for gamma-ray burst progenitor mass}
\author[J.~Larsson et al.]{
Josefin Larsson$^{1,3}$, Andrew J.~Levan$^{2,3,4}$, Melvyn B.~Davies$^3$,
Andrew S. Fruchter$^5$\\
$^1$Institute of Astronomy, Madingley Road, Cambridge CB3 0HA, UK\\
$^2$Centre for Astrophysics Research, University of Hertfordshire, 
College Lane, Hatfield AL10 9AB, UK\\
$^3$Lund Observatory, Box 43, SE--221 00 Lund, Sweden.\\
$^4$Department of Physics, University of Warwick, Coventry, CV4 7AL, UK \\
$^5$Space Telescope Science Institute, 3700 San Martin Dr., 
Baltimore, MD 21218, USA\\
}

\date{Accepted 2007 January 18}

\pagerange{\pageref{firstpage}--\pageref{lastpage}}
\pubyear{2007}

\begin{document}

\maketitle

\label{firstpage}

\begin{abstract}
  Recent comparative observations of long duration gamma-ray bursts
  (LGRBs) and core collapse supernovae (cc SN) host galaxies
  demonstrate that these two, highly energetic transient events are
  distributed very differently upon their hosts. LGRBs are much more
  concentrated on their host galaxy light than cc SN. Here we explore
  the suggestion that this differing distribution reflects different
  progenitor masses for LGRBs and cc SN. Using a simple model we show
  that, assuming cc SN arise from stars with main sequence masses $>$
  8 M$_{\odot}$, GRBs are likely to arise from stars with initial
  masses $>$ 20 M$_{\odot}$.  This difference can naturally be
  explained by the requirement that stars which create a LGRB must
  also create a black hole.
\end{abstract}


\section{Introduction}
Long duration gamma-ray bursts (GRBs) originate in hydrogen deficient
core collapse supernovae (SN Ic [Woosley 1993; Hjorth et al. 2003;
Stanek et al. 2003]).  These supernovae differ from the bulk of the
core collapse population by showing no discernible hydrogen or helium
lines, and, often exhibiting very high velocities ($\sim 30000$ km
s$^{-1}$).  The most likely candidate progenitor systems are thus wolf
rayet stars which have lost their hydrogen envelopes via binary
interactions, stellar winds or, possibly via complete mixing on the
main sequence (e.g. Izzard et al. 2004; Podsiadlowski et al. 2004a;
Yoon \& Langer 2005; Woosley \& Heger 2006). Absorption spectra of GRB
afterglows support this picture, with some bursts exhibiting several
different absorption systems with velocity shifts of several hundred
km/s, possibly wolf-rayet shells from the progenitor wind (e.g.
Starling et al. 2004; van Marle et al. 2005).

However, measuring the properties of the progenitor star is complex.
Although in a few, nearby cases it is possible to infer the properties
of the star prior to core collapse by detailed modelling
of the supernova spectrum, the paucity of nearby events limits this
sample.  An alternative means of understanding the nature of GRB
progenitors is via the study of their galactic environments. Clearly
any study conducted after the event cannot contain the progenitor star
itself, however the environment should be indicative of the star
formation which was occurring at the time of the GRB.  Recently
Fruchter et al.  (2006) have conducted a survey of the galactic
environments of both long duration GRBs and core collapse SNe (i.e.
all types of core collapse events, including SN II, Ib and Ic). These
results demonstrate that GRBs are highly concentrated on their host
light, significantly more so than the core collapse supernova
population.  Fruchter et al. (2006) further suggest that this can be
explained as being due to the GRBs originating in the most massive
stars, which, upon core collapse form black holes rather than neutron
stars. Here we further explore this possibility and attempt to derive
plausible limits on the progenitor lifetime and mass based on the
observed distributions of cc SNe and GRBs upon their host light. Using
a simple model, motivated by the distributions of young star clusters
in a local starburst galaxy, we explore the expected distributions of
stars of different masses upon their host galaxies and compare these
to the observed distributions from Fruchter et al. (2006). Our results
demonstrate that for plausible models more massive stars are {\it
  always} more concentrated on their host light than lower mass stars.
Further, given that supernovae originate from stars with initial
masses $>$ 8 M$_{\odot}$, we find that the observed distributions of
GRBs on their host galaxies can naturally be explained by progenitors
with initial masses in excess of 20 M$_{\odot}$.

\section{Model}
\label{model}
\subsection{A local starburst galaxy as a template}\label{template}
GRB host galaxies at high redshift are starburst galaxies, with high
specific star formation rates (i.e. star formation rates per unit mass
- e.g. Christensen et al. 2004).  A natural local analogue for such a
galaxy is NGC 4038/39 -- the Antennae, and here we use it as a template for
constructing a simple model of a GRB host.

NGC 4038/39 have been studied in detail with the {\it Hubble Space
  Telescope} ({\it HST}) and the young star clusters have been
identified on the basis of H$\alpha$ imaging (Whitmore \& Schweizer
1995; Whitmore et al. 1999). Furthermore the age of young star
clusters has been determined on the basis of their multicolour
properties compared to the expected synthetic colours of clusters at
different ages (Fall et al. 2005). The luminosity function and surface
density of clusters on NGC 4038/39 is comparable to that seen in other
local star forming galaxies of varying morphology (e.g. M51 or even
the LMC, Gieles et al.  2006), indicating that it is a reasonable
template.  Using this well defined sample of clusters it is possible
to examine where they lie on their galaxy light as a function of, for
example, cluster age and luminosity.

Of course NGC 4038/39 lies only $\sim 20$ Mpc distant, as such the
resolution of the observations are much higher than is possible for
GRB host galaxies at $z=1$. Thus we resampled the observations of NGC
4038/39 as they would appear at $z=1$ (using a $\Lambda$CDM cosmology
with $\Omega_{\Lambda} =0.73, \Omega_{M} = 0.27, H_0 = 71$ km s$^{-1}$
Mpc$^{-1}$).  We used the u-band (F330W) WFPC2 observations of NGC
4039, which broadly corresponds to the rest-frame wavelength observed with the
F606W filter (the most common filter used in Fruchter et al. 2006) at
$z=1$. We then use the total pixel distribution and the total cluster
distribution to determine the key parameters of the model (described
below).  Subsequently our model galaxy was set up based on these
observations so that it reasonably represents a GRB host galaxy at
$z=1$.


\subsection{Parameters and implementation of the model}
\label{parameters}

A common approach when studying the environments of astronomical
objects is to determine the distance of the object in question from
the centroid of its host galaxy's light.  This method however provides
limited information when studying GRBs and SNe as many of their hosts
are irregular galaxies with more than one bright component.  Fruchter
et al. (2006) therefore developed a method which is independent of
galaxy morphology. In their survey the position of a GRB or a SN was
determined by sorting all of the pixels of the host galaxy from
faintest to brightest, and asking what fraction of the total light is
contained in pixels fainter than or equal to the pixel containing the
explosion.

Our aim here is to set up a simple model which can be directly
compared with the observational results obtained by Fruchter et al.
(2006). We therefore define the properties of our model in terms of
the contents of each pixel in a galaxy consisting of 500 pixels (the
mean number of pixels in the observational sample of GRB hosts).
Specifically each pixel is given a number of (or no) young clusters as
well as light from old clusters and low-mass field stars.  The light
from old clusters and low-mass stars will from here on be referred to
as background light. Because of the short lifetimes of GRB and SN
progenitors we take the background light to be a constant in our
model.  The young clusters are created according to the age and mass
distributions described below and their luminosities evolve with time
as individual stars end their lives. The key parameters of our galaxy
model are:

\begin{itemize}
\item
The surface density of clusters.
\item
The distribution of cluster masses.
\item
The distribution of cluster ages.
\item
The distribution of background light.
\item
The distribution of clusters on the background light.
\end{itemize}

\begin{center}
\begin{figure}
\resizebox{75mm}{!}{\includegraphics{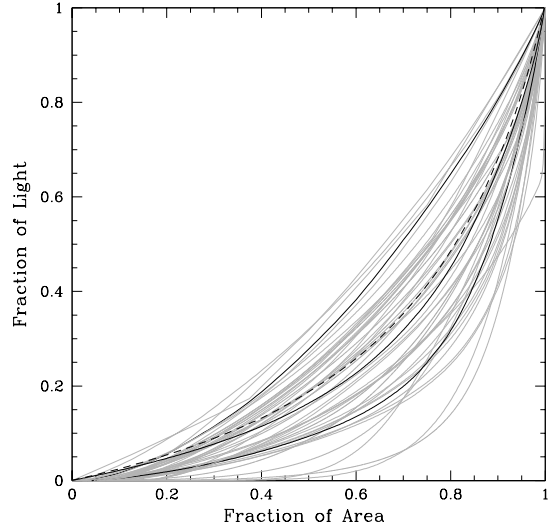}}
\caption{Light profiles for the observed GRB and SN hosts (in grey)
  together with the profile of the background light (dashed black
  line) and total light (thick solid black line) in our model. The
  extreme points of the distributions for the SN hosts are shown as
  thin solid black lines.} \label{hosts}
\end{figure}
\end{center}
The first two properties were taken directly from the observations of
NGC 4038/39 at $z\sim1$ as described in section \ref{template}. The
surface density of clusters expressed in terms of number of clusters
per pixel is $\sim 0.15$, although this is far from uniform across the
galaxy.  We use the observed distribution of cluster masses, which
follows $dN/dM_{\rm{cl}}\propto M_{\rm{cl}}^{-2}$, with
$M_{\rm{cl,min}}=4\cdot 10^4 \ M_{\odot}$ and $M_{\rm{cl,max}}=10^6\
M_{\odot}$, where $M_{\rm{cl}}$ is the cluster mass.  The age
distribution of clusters was taken from Fall et al. (2005) and follows
$dN/d\tau \propto \tau^{-1}$, where $\tau$ is the cluster age. In
order to mimic an (almost) instantaneous burst of star formation,
clusters are created in the model according to this distribution over
a period of $10^7$ years.

To address the issue of the distribution of background light we
investigated the light profiles of all the GRB and SN hosts in the
observational sample. These are plotted in grey in Fig. \ref{hosts}.
The light distributions for the two types of hosts are very similar
although the distributions of the SN hosts fall in a somewhat narrower
range than those of the GRB hosts (extreme points for the SN hosts are
shown as thin black lines in the figure).  This discrepancy is likely
to at least in part be due to the smaller number of SN hosts present
in the observational sample (the sample contains 16 SN hosts and 32
GRB hosts).

Because the total background light is larger than the total light from
young clusters (a factor of 6--8 for NGC 4038/39) we can use the
total light profiles of Fig. \ref{hosts}. to obtain an expression for
the distribution of background light in our model. The distribution we
adopted lies roughly in the middle of the observed distributions
(black dashed line in Fig. \ref{hosts}) and is given by
$dN/dL_{\rm{pix}} \propto L_{\rm{pix}}^{-1.5}$ with
$L_{\rm{pix,max}}/L_{\rm{pix,min}}=20$, where $L_{\rm{pix}}$ is the
luminosity of a pixel. The solid black line shows the distribution of
total light in the model after cluster light has been added as
described below.

\begin{center}
\begin{figure}
\resizebox{75mm}{!}{\includegraphics{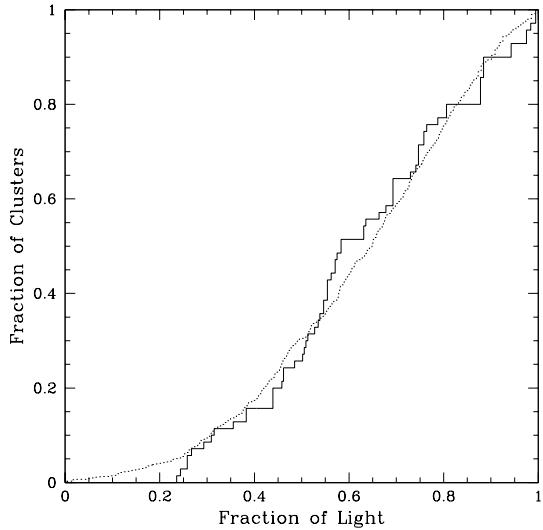}}
\caption{Distribution of clusters on the total light for NGC 4038/39
  at $z=1$ (dotted line) and our model (solid line). The distribution
  for the model was obtained 14 Myr after the first clusters were
  created. Note, that while the clusters within NGC 4038/39 of course
  show a range of ages the mean age (Fall et al. 2005) is
  comparable to that plotted for our model.}\label{clusters}
\end{figure}
\end{center}

The distribution of clusters on the background light is slightly more
complicated than the other items since it is hard to observationally
separate the cluster light from background light in individual pixels.
We do however see a clear correlation between the number of clusters
and the total light for the pixels in NGC 4038/39. Since background
light makes up most of the total light we conclude from this that
young, massive clusters are more likely to be found in pixels with a
higher amount of background light. To account for this in our model
we developed a correlation method in which the probability of a given
pixel containing clusters increases with the amount of background
light in that pixel. Fig. \ref{clusters} shows the resulting
distribution of clusters on the light 14 Myr after the first clusters
were created (solid line). The figure also shows the distribution of
the clusters in NGC 4038/39 as it would look at $z=1$ (dotted line).
The two distributions show excellent agreement. Further the
distribution of clusters, in terms of number of clusters per unit
pixel luminosity, also shows excellent agreement with the observations
of NGC 4038/39 as it would appear at $z=1$.  The cluster distribution
of course evolves with time but it is encouraging that the model
resembles NGC 4038/39 at a time when several GRBs are occurring.

In order to identify GRBs and SNe and follow the evolution of the
young clusters, each cluster was populated with stars drawn from a
Salpeter IMF ($dN/dM \propto M^{-2.35}$) 
during a period of $10^{6}$ years. The luminosity of a
star was taken to go as $L_{\ast}\propto M_{\ast}^3$ (to approximate
blue light) and the stellar lifetime was approximated by the main
sequence life time according to $T_{\ast} \propto 4\cdot 10^6 \cdot
(100/M_{\ast})$~years. These assumptions are somewhat simplistic but
provide good agreement with results from more complete stellar
evolution calculations (e.g. Pols et al. 1995; Hurley, Pols \& Tout
2000) and are sufficient for our purposes here.

With this setup, the total luminosity of our galaxy right after all
the clusters have been created is about $3\cdot 10^{10}\ L_{\odot}$ and the
total number of young clusters is around 70. We note that the total
luminosity of our model galaxy is higher than for a typical GRB host
(these are typically around $10^{10}\ L_{\odot}$), but the properties
of our model scale with luminosity and our results are therefore not
affected by this.

In each run of the program, roughly 500 stars which are more massive
than a specified minimum progenitor mass are randomly selected, the
position on the light for each selected object is calculated at the
end of its lifetime, and the cumulative distribution showing the
fraction of objects as a function of fraction of light is produced.
Because the model galaxy evolves with time the galaxy looks somewhat
different for every recorded SN or GRB, and we assume that these
differences account, to first order, for the differences between the
observed host galaxies.

\section{Results}

\begin{center}
\begin{figure}
\resizebox{75mm}{!}{\includegraphics{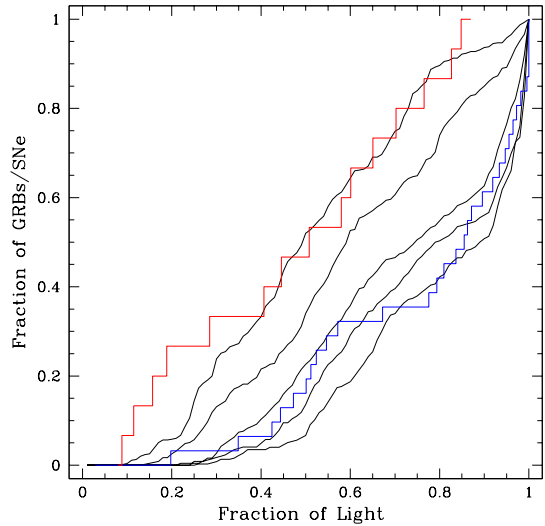}}
\caption{Fraction of objects plotted against fraction of light for
  observed SNe (red) and GRBs (blue) together with the results from
  our model (black lines). Black lines from top to bottom correspond
  to minimum progenitor masses of 8, 20, 40, 60, and 80
  $M_{\odot}$.}\label{grbsn}
\end{figure}
\end{center}

\begin{center}
\begin{figure}
\resizebox{75mm}{!}{\includegraphics{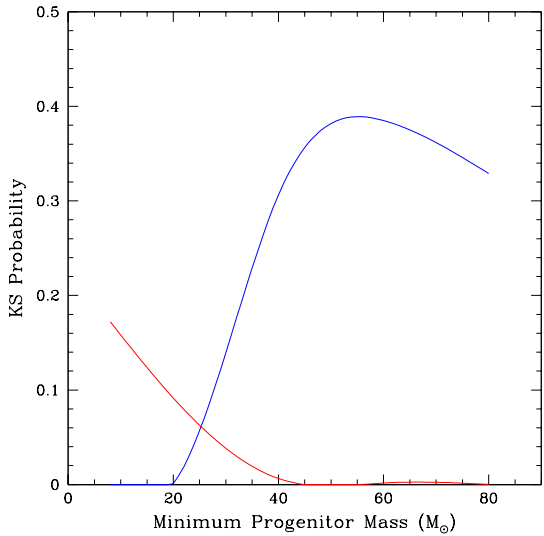}}
\caption{KS-probabilities of our model results following the observed
  SN (red) and GRB (blue) distributions. The probabilities are plotted
  as a function of the minimum progenitor mass and were calculated for
  masses of 8, 20, 40, 60, and 80 $M_{\odot}$. A spline has been
  fitted through the data points.}\label{ks1}
\end{figure}
\end{center}

Using the parameters described in the previous section we performed
runs for minimum progenitor masses of 8, 20, 40, 60, and 80
$M_{\odot}$. The results are shown as black lines in Fig. \ref{grbsn}
together with the observed distributions of SNe (in red) and GRBs (in
blue) from Fruchter et al. (2006).
The model distributions for all masses were KS-tested against the
observed SN and GRB distributions and the resulting probabilities are
shown as a function of mass in Fig. \ref{ks1}.  While the probability
of following the SN distribution decreases with increasing mass, the
likelihood of following the observed GRB distribution increases
rapidly from 8 to 40 $M_{\odot}$ and then flattens out, reaching a
weak maximum around 60 $M_{\odot}$. The shapes of the two probability
functions look the same for all realisations of the model, although
the peak probabilities can change by about 0.1 between different runs.
These results strongly suggest that GRB progenitors are significantly
more massive than SN progenitors.

\section{Discussion}

\subsection{Robustness and limitations of the model}

In this section we address the robustness of our results by
considering the errors on the observed distributions as well as the
uncertainties and limitations of our model.

An issue requiring discussion is how one matches our theoretical
model to the observational data. Clearly the observations contain
various measurement errors, whereas errors within the model
are contained within the assumption which are made. 
The observational errors to consider are those on the photometry
(i.e. the error on the value of the pixel containing the GRB or SN) and
those on the astrometry (i.e. knowledge of the location of the
transient on its host galaxy). The latter is normally very small 
(though see Fruchter et al. 2006 for a more complete discussion), while
the former depends largely on the value of the pixel in question, 
bright pixels have markedly smaller measurement errors, while 
fainter pixels can change their position (as a function of host
galaxy light) by up to $\sim 10$\% based on the typical 
1$\sigma$ noise within a pixel. However, we expect that this 
effect will average out over the larger sample.

Additionally, observations at high redshift do not reveal the full
optical extent of a given galaxy, since light contained within pixels
of low surface brightness is not detected above the sky level.
Although the majority of the light is concentrated in brighter regions
the faintest pixels (typically corresponding to a few percent of the
light at $z \sim 1$) are not detected.  To mimic this effect we
employed a surface brightness cut upon our models, removing the
faintest pixels containing about 5\% of the light, although we note
that qualitatively our results are not strongly dependent on the
effects of this cut, since the majority of the light is contained in
brighter pixels.

The results from our model show that the most massive stars are
significantly more concentrated on their host galaxy light than the
$\sim 8 M_{\odot}$ stars which give rise to the bulk of the cc SN.
This is simply a consequence of the different lifetimes of stars of
different masses; the most massive stars are found in bright, young
clusters which can provide the peak of the light of a galaxy, while
most of the SN occur when their clusters are fainter and therefore
less likely to be in the brightest parts of a galaxy.  

The exact positions on the light for GRBs/SNe with different
progenitor masses of course depend on the parameters of our model.
Because the model contains numerous free parameters with relatively
weak constraints on their range and correlation from direct
observations, we have chosen not to do detailed simulations covering
all of parameter space, but simply to show that we can get good
agreement with observations for a reasonable set of parameters. In
order to investigate the robustness of our results we however
performed several runs varying each of the key parameters listed in
section \ref{model} while keeping the rest of the model fixed.  We
found that the most important parameters are the level of background
light, the distribution of clusters upon this background, and the
number of young clusters.

In order to investigate the effect of varying the level of background
light we performed runs with a total background ranging from 1/3 to 10
times the background of our standard model. The lower limit is set by
requiring that the total cluster light never exceeds the background
light in our model. We note that our analysis of the starburst galaxy
NGC 4038/39 finds a background--to--cluster light ratio of around
6--8, and that therefore extremely unusual conditions would be needed
to arrive at our lower limit. The upper limit corresponds to what
would be expected in early type galaxies which contain relatively few
supernovae and are equally not expected within our sample of GRB or SN
hosts.

Decreasing the amount of background light in the model makes the
contribution from cluster light more important and all progenitors
therefore become more concentrated on the host light. For the lowest
background the 8~$M_{\odot}$ progenitors fall between the observed SN
and GRB distributions. Increasing the background has the opposite
effect and for the highest background all the progenitors are less
concentrated on the light than the observed SNe. Because the
background completely dominates the total light distribution close to
our upper limit, the distributions for different progenitor masses
also move closer together. More massive progenitors are however always
more concentrated on the light than lower mass ones.

In this case one may wonder if it is possible for the GRB distribution
to be explained by differing background to cluster light ratios in the
host galaxies. However, as GRB hosts typically have a high specific
star formation rates (i.e. high star formation rates per unit total
luminosity) we would expect that GRB hosts would have higher cluster
to background light ratios. However, we note that even in the extreme
case of equal background and cluster light (which is unreasonable in
essentially all galaxies) the distribution of SN progenitors remains
{\em less} concentrated than is observed for GRBs.

In the runs just described we varied the amount of background light
while keeping the shape of the distribution constant. As described in
section \ref{parameters}, this shape was chosen as the median of the
light distributions of all the hosts seen in Fig. \ref{hosts}.  To
check whether this simplification has any effect on the result we
compared the result obtained when using only the median to the result
obtained by averaging the results for different light distributions
drawn from Fig.  \ref{hosts}. We found that the results are indeed
very similar.
  
As mentioned in section \ref{parameters} the distribution of clusters
in NGC~4038/39 suggests that young clusters are more likely to be
found in pixels with a higher level of background light.  Since it is
hard to put observational constraints on this correlation for
different types of galaxies we simply note that both no correlation
and maximal correlation are unphysical: young clusters are always
found in bright regions of galaxies and there is simply not enough
space to place all of the clusters in the very brightest pixels. We
therefore performed runs varying the correlation between these two
extremes.  When little correlation is present progenitors of all
masses are less concentrated on the host light and the distribution
for 8~$M_{\odot}$ progenitors is pushed above the observed SN
distribution. Increasing the correlation has the opposite effect but
the distribution for 8~$M_{\odot}$ progenitors still remains well
above the observed GRB distribution. For all correlations more massive
stars are always more concentrated on the light than lower mass ones.

As a side note we also point out that the correlation is degenerate
with the level of background light; a similar result can be obtained
using a high background together with a high correlation as with a
lower background together with a low correlation.

The last parameter which was seen to significantly affect the results
is the surface density of young clusters.  Since the surface density
of clusters on NGC 4038/39 is comparable to that seen in other local
star forming galaxies (Giels et al.  2006) we simply vary the number
of clusters in the model between 1/10 and 10 times this typical value.
When fewer young clusters are included the progenitors are more
concentrated on the light as there are more low luminosity pixels
present. Including a larger number of clusters makes the progenitors
less concentrated on the light.  As in the case of varying the
background and correlation we however find that more massive
progenitors are always more concentrated on the light than lower mass
ones, and that the 8~$M_{\odot}$ progenitors are always well above the
observed GRB distribution.  Indeed, as we might expect the surface
density of clusters to be higher in GRB hosts (because of their high
specific star formation rates) this effect would bias the observed
results in the {\em opposite} direction from that observed (i.e. it
would typically make GRB progenitors seem less concentrated on their
hosts).

In summary our results are robust in the following important aspects:

\begin{itemize}
\item For all plausible models more massive progenitors are
  \emph{always} more concentrated on their host galaxy light than
  lower mass progenitors.
  \\
\item We found no reasonable parameters for which a progenitor mass of
  $8\ M_{\odot}$ was close to following the observed GRB distribution,
  indicating that the GRB progenitors have significantly higher
  masses.\\
\item In models where progenitors with a minimum mass of 8 $M_{\odot}$
  follow the observed SN distribution, we always find that the GRB
  progenitors have to be more massive than 20 $M_{\odot}$.
\end{itemize}
Based on these results we conclude that the observed locations of SNe and
GRBs on the light of their host galaxies can be explained if the GRB
progenitors are significantly more massive than the SN progenitors.

Since the lifetime of a star is clearly dependent on its mass we can
also place a limit on the lifetime of the stars forming the GRBs. The
observed light comes from the ensemble of massive stars around the
progenitor of the GRB, and it is the age of these which effectively
sets the distributions seen in Figure 3. It is plausible that, as
unusual stars, GRB progenitors might follow different evolutionary
paths and thus have different lifetimes. For example, under the models
of Yoon \& Langer (2005) and Woosley \& Heger (2006) rapidly rotating
massive stars undergo complete mixing and have longer lifetimes than
normal main sequence stars (with lower angular momentum) of the same
mass. In this sense the distributions may more accurately set the age
of the population rather than the mass of the progenitor.  The
distributions shown in Figure 3 are well reproduced by a population
with $t_{SN} < 50$~Myr and $t_{GRB} < 20$~Myr. We note that these
values differ from those obtained via detailed stellar evolution
modelling and that most models predict shorter life times (e.g.
Schaller et al. 1992). This discrepancy is due to the relatively
simple treatment of the main sequence lifetimes within our code.

In deriving these results we have, of course, assumed the same basic
model for GRB and SN hosts, in terms of the expected relative
distribution of clusters upon them. In truth this is poorly known,
although it is clear that the different host galaxies differ
significantly morphologically, with 50\% of SN hosts being spiral,
compared to only 7\% of the GRB hosts in the same redshift range.
Typically the star formation is likely to be more intense in the GRB
host galaxies than in the SN sub-sample. So, in the language of our
model the background to cluster light ratio will be lower in GRB host
galaxies.  This is to some extent taken into account in our model as
the most massive stars end their lives while the majority of the stars
in the cluster are still on the main sequence (i.e. while the cluster
is at its maximum luminosity), whereas most $8 M_{\odot}$ stars will
explode as SNe when much of the cluster light has disappeared.  In
practice, measuring the distribution of individual clusters on GRB and
SN host galaxies will be impossible for the foreseeable future, and we
can not assess how well our model accounts for these plausible
differences between the two types of hosts.  We therefore caution
against drawing strong quantitative conclusions on GRB progenitor mass
from our models but note that it must {\it always} be significantly
larger than the SN progenitor mass.

\subsection{Implications of the results}

A natural explanation for the preference of GRBs to occur from more
massive stars is that most cc SN create neutron stars, while GRBs are
likely to originate from black holes. Indeed, models imply that the
dividing line between NS and BH creating supernovae occurs at around
25 M$_{\odot}$ (Heger et al. 2003), in general agreement with the
limits which we derive here.

It is important to note that while we here derive a limit on the
GRB progenitor mass of $\ga$ 20 M$_{\odot}$ this does not
imply that all stars above this mass will create GRBs. Indeed the rate of
formation of stars with main sequence masses of $>20$ M$_{\odot}$ 
exceeds the GRB rate by roughly two orders of magnitude. 
Additional effects clearly reduce the rate of formation of GRBs.
The most likely of these are metallicity, rotation and
the presence of a hydrogen envelope. The
comparative study of Fruchter et al. (2006) demonstrates that
typically GRB hosts are less luminous and smaller than 
those of cc SN. This implies that GRB host galaxies will
exhibit a lower global metallicity, while SN hosts are more 
luminous and likely have higher metallicities.

An additional constraint for GRB production comes from 
rotation. GRB production is thought to require the formation 
of a torus about the nascent black hole. These discs can only
be formed in rapidly rotating stars, and therefore further
limit the fraction of massive stars which can create GRBs.
The majority of single stars rotate too slowly 
on the main sequence for torus formation, and only 
a small fraction are in binaries with sufficiently small
separations for tidal locking to create stars with
sufficient rotation for torus formation (Izzard et al. 2004;
Podsiadlowski et al. 2004a; Levan, Davies \& King 2006).

Finally, it should be noted that all of the SN spectroscopically
associated with GRBs are of the type Ic. The lack of
discernible hydrogen (or helium) lines in these spectra implies
that the progenitor stars have lost their hydrogen (and helium)
envelopes prior to core collapse. Therefore, even very
massive stars cannot be considered candidates for
GRB production if they retain significant hydrogen atmospheres.

In deriving the results above we have assumed that all stars with M
$>$ 8 M$_{\odot}$ create core collapse supernovae, and have not
attempted to differentiate between different subtypes of these (e.g.
SN II (H-rich) and SN Ib/c (H-poor)). This is reasonable since SN II's
dominate the {\it observed} rate. However, there are reasons for
believing that the brightness of a supernova, and therefore its
detectability at high redshift, may not be independent of the
progenitor mass (provided it is greater than 8 M$_{\odot}$.  For
example, stars in the range 8-10 M$_{\odot}$ may undergo electron
capture supernovae (e.g. Nomoto 1987; Podsiadlowski et al. 2004b),
while stars with masses in excess of $\sim$ 25 M${\odot}$ might create
black holes either by direct collapse or fallback, but without a
bright supernova (e.g. Heger et al. 2003). Both of these may create
supernovae with faint optical emission, and might not be represented,
even in deep optical surveys. We tested this effect by creating a
model in which supernovae were drawn exclusively from the masses in
the range $10 < M_{SN} < 25$.  Although this slightly alters the shape
of the distribution seen in Fig. 3 it still provides an excellent
agreement to the supernova distribution ($P_{KS} = 0.23$, compared to
$P_{KS} = 0.17$ for the $M_{SN} > 8$ distribution. )

Clearly as GRBs are rare events it is possible, or even likely, that
the progenitors follow exotic pathways to their production.  These
pathways may plausibly involve binary interactions or even collisions
(which can build up even more massive stars). As the number of
interactions scales roughly as the 3/2 power of the mass of the
cluster (e.g. Davies, Piotto \& De Angeli 2004) more massive (and
hence brighter) clusters might harbour more GRBs. We note that simply
picking GRBs where the probability of a GRB occurring is proportional
to the mass of the cluster does not accurately reproduce the
observations.  It may well be that other parameters, such as cluster
core densities, are also important. However a full investigation of
these is beyond the scope of this paper.

\section{Summary}

An observational study by Fruchter et al. (2006) showed that long
duration GRBs are significantly more concentrated on their host galaxy
light than core collapse SNe. In this paper we have used this result
in an attempt to put constraints on the mass of GRB progenitors.  In
order to construct a simple model of a typical GRB host we used the
properties of the local starburst galaxy NGC 4038/39 as it would appear
at a redshift of 1. We then specified different minimum masses of
GRB/SN progenitors and studied their locations on the light of our
model galaxy.  We showed that the observed locations of SNe
and GRBs on the light of their host galaxies can be explained if the
GRB progenitors are significantly more massive than the SN
progenitors. The exact value of the minimum GRB progenitor mass
depends on the parameters of our model, but for a reasonable set of
parameters the minimum progenitor mass was found to be significantly
higher than $20 M_{\odot}$.

\section*{Acknowledgements}
JL thanks Corpus Christi College, the Isaac Newton Trust and PPARC for
studentship awards. AJL is grateful for support from a PPARC
postdoctoral fellowship, and thanks the Swedish Institute for support
while visiting Lund Observatory. MBD is a Royal Swedish Academy
Research Fellow supported by a grant from the Knut and Alice
Wallenberg Foundation.

{}

\end{document}